\documentclass[preprint,12pt,authoryear]{elsarticle}
\usepackage{url}
\usepackage{amssymb}
\usepackage{amsmath}
\usepackage{graphicx}
\usepackage[inline]{enumitem}
\usepackage{booktabs} 
\journal{Computers in Human Behavior}

\begin{document}
\begin{frontmatter}

\title{Affective-CARA: A Knowledge Graph–Driven Framework for Culturally Adaptive Emotional Intelligence in HCI} 

\author[inst1]{Nirodya Pussadeniya}
\ead{en95142@sjp.ac.lk}

\author[inst2]{Bahareh Nakisa}
\ead{bahar.nakisa@deakin.edu.au}

\author[inst3]{Mohmmad Naim Rastgoo}
\ead{naim.rastgoo@monash.edu}

\address[inst1]{University of Sri Jayewardenepura, 
                Colombo, 
                Sri Lanka}
\address[inst2]{School of Information Technology, 
                Faculty of Science, Engineering and Built Environment, 
                Deakin University, 
                Geelong, Vic, 
                Australia}

\address[inst3]{Faculty of Information Technology,                                  Monash University, 
                Australia}

\begin{abstract}

Culturally adaptive emotional responses remain a critical challenge in affective computing. This paper introduces Affective-CARA, an agentic framework designed to enhance user–agent interactions by integrating a Cultural Emotion Knowledge Graph (derived from StereoKG) with Valence, Arousal, and Dominance annotations, culture-specific data, and cross-cultural checks to minimize bias. A Gradient-Based Reward Policy Optimization mechanism further refines responses according to cultural alignment, affective appropriateness, and iterative user feedback.

A Cultural-Aware Response Mediator coordinates knowledge retrieval, reinforcement learning updates, and historical data fusion. By merging real-time user input with past emotional states and cultural insights, Affective-CARA delivers narratives that are deeply personalized and sensitive to diverse cultural norms. Evaluations on AffectNet, SEMAINE DB, and MERD confirm that the framework consistently outperforms baseline models in sentiment alignment, cultural adaptation, and narrative quality.

Affective-CARA achieved a Cultural Semantic Density of 9.32 out of 10 and lowered cultural representation bias by 61\% (KL-Divergence: 0.28), demonstrating robust performance in generating ethical, adaptive responses. These findings suggest the potential for more inclusive and empathetic interactions, making Affective-CARA an avenue for fostering culturally grounded user experiences across domains such as cross-cultural communication, mental health support, and education.
\end{abstract}

\begin{highlights}
\item Proposes a novel framework for real-time, culturally adaptive affective responses
\item Hyperbolic culture–emotion graph reduces bias by 61\% vs. SenticNet-GPT baseline
\item Achieves +39\% emotional appropriateness and 9.32 CSD, SOTA in cultural adaptation
\item Proposes a new model for culture-aware response adaptation based on GRPO RLHF
\item CARM meta-controller unifies recognition, cultural reasoning and narrative generation
\end{highlights}


\end{frontmatter}

\section{Introduction}
Affective computing systems designed to interpret and respond to human emotions has experienced rapid advancements, particularly driven by deep learning techniques and large-scale language models \citep{ERCsurvey2024}. 

Decades of cross-cultural psychology show that the way people express, perceive, and regulate emotion is strongly culture-bound.  Concepts such as display rules and the ``in-group advantage’’ demonstrate that identical facial or verbal cues can convey different affective meanings across societies, and observers decode emotions more accurately within their own culture group.  Consequently, technologies that ignore culture risk misclassifying users’ affect or delivering responses perceived as insensitive.  These risks are amplified in sensitive domains such as mental-health support, education, and global customer service, where user trust is paramount.

Despite recent progress, effectively addressing cultural variability in emotional expression remains a critical, yet unresolved challenge. Recent research underscores significant performance deterioration (approximately 20-40\%) when affective computing models trained primarily on Western-centric datasets are deployed in culturally diverse contexts \citep{PeiAffectiveComp2024, HuAffective2025}. These findings highlight persistent limitations in current emotion recognition and response generation frameworks, notably their inability to generalize well across different cultures due to inherently biased data and inadequate adaptive mechanisms \citep{BarthetACII2022, EmoTransKG2024}.

Most existing systems either (i)~focus solely on improving cross-cultural  recognition accuracy, or (ii)~build static cultural profiles that cannot adapt in real time.  Critically, they  rarely close the loop that is, they do not convert cultural knowledge into  adaptive emotional responses.  No prior work unifies a structured representation of culture with a learning mechanism that tunes the agent’s behaviour on-the-fly.  Addressing this gap is essential for culturally inclusive, empathetic HCI.

Current affective systems predominantly rely on universal-emotion taxonomies and static user-profiling methods \citep{ERCsurvey2024}. Although recent integration of knowledge graphs with large language models (LLMs) has demonstrated improvements in emotion understanding and response generation \citep{ECoK2024, sabour-etal-2024-emobench}, these approaches generally utilize Euclidean embeddings. Such flat representations fail to capture the inherent hierarchical and nuanced nature of cultural-emotional relationships, limiting their cross-cultural adaptability and often resulting in stereotyped or contextually inappropriate responses \citep{LI2023104503, ECAI2024}.

Furthermore, contemporary reinforcement learning approaches in affective computing typically neglect explicit cultural reward mechanisms, limiting their adaptive capabilities in culturally nuanced environments \citep{DannICML2023, PervezMetaverse2024}. Consequently, existing methods inadequately address users' dynamic emotional needs, perpetuating biases and underscoring the necessity for culturally responsive frameworks \citep{GilbertRewardReports2023}.

To address these critical gaps, we introduce Affective–CARA (Culturally Adaptive Response Agent), an innovative framework specifically designed to generate culturally adaptive, emotionally intelligent responses. Affective–CARA uniquely integrates a hyperbolic Cultural Emotion Knowledge Graph enhanced with Valence-Arousal-Dominance (VAD) annotations and cultural alignment checks. This structure effectively captures hierarchical cultural relationships, significantly reducing representational bias. Moreover, our approach employs Gradient-based Reward Policy Optimization (GRPO) tailored explicitly to cultural appropriateness and emotional intelligence signals, allowing dynamic adaptation in response to real-time user feedback. Finally, the Cultural-Aware Response Mediator (CARM) systematically combines real-time inputs with historical emotional states and cultural insights, enabling deeply personalized narrative generation.

Specifically, our contributions are threefold. \begin{itemize}[]
  \item  CEKG: a hyperbolic knowledge graph that encodes hierarchical culture–emotion norms and lowers representation bias by 61\,\% (vs. a SenticNet-GPT baseline, p < .01).
  \item  GRPO: a novel RL strategy that optimises responses for cultural appropriateness and emotional coherence in real time.
  \item  CARM: a meta-controller that unifies recognition, cultural reasoning, and narrative generation, enabling end-to-end adaptive emotional interaction.
\end{itemize}

Comprehensive evaluations conducted across culturally diverse datasets including MERD \citep{OVMER2024}, EmpatheticDialogues \citep{KBDG2022}, and SEMAINE \citep{SEMAINE2012} demonstrate that Affective–CARA consistently outperforms state-of-the-art methods on metrics of cultural semantic richness, bias reduction, emotional appropriateness, and narrative quality.

\section{Related Work}

The field of affective computing originated with vision for emotionally aware machines, launching early efforts that modeled emotion generically and overlooked individual or cultural differences. Later studies demonstrated that emotional expression and perception are highly culture-dependent, with models trained in one cultural context often performing poorly elsewhere \citep{AfzalAffectiveSurvey2023}. While foundational agent models such as \citep{MascarenhasSocial2013} incorporated static cultural profiles using Hofstede’s dimensions, they could not learn or generalize beyond pre-defined cultural stereotypes.

Culture profoundly shapes how emotions are expressed and perceived, yet traditional NLP benchmarks often ignore these nuances. Belay et al. (2025) introduce CuLEmo, a benchmark for culture-aware emotion understanding across six languages, explicitly designed to require nuanced cultural reasoning \citep{BelayCuLEMO2025}. Their findings show that emotion concepts and model performance vary significantly with cultural context, underscoring that “emotion conceptualizations vary significantly across languages and cultures”. Similarly, \citep{DeasMASIVE2024} present MASIVE, a dataset of over 1,000 unique affective states in English and Spanish, to move beyond a fixed set of basic emotions. They note that basic emotion categories (e.g. anger, joy) are insufficient for textual data, as “culture, language, and dialect can influence how particular emotions are interpreted”. Their models show improved performance when trained on native, culturally-specific data, highlighting the importance of culturally grounded emotional lexicons.

Large Language Models (LLMs) trained on predominantly Western data often exhibit cultural bias, favoring certain norms or dialects \citep{LiCulturePark2024}. This can lead to harm and unfair performance for under-represented groups. For example, \citep{HofmannNature2024} demonstrate that an AI system produced “covertly racist decisions” based solely on a user's dialect. In a similar vein, \citep{DeasAALBias2023} evaluate GPT-style models on African American English and find significant performance gaps compared to mainstream English, indicating bias and lack of understanding of that dialect’s linguistic features. These studies emphasize the need for culturally inclusive models. Indeed, \citep{CaoCulturalDialogue2024} probed ChatGPT using Hofstede’s cultural values survey and found a “distinct disparity between the system and human society” in value alignment. \citep{arora-etal-2023-probing} likewise report that pretrained language models encode different value judgments across cultures. Such analyses confirm that current LLMs are not culture-neutral; they carry latent cultural assumptions that can misalign with users from diverse backgrounds. Researchers have called for pluralistic alignment models that can flexibly accommodate multiple cultural value systems. \citep{SahaMetaCultural2025} even argue that beyond factual cultural knowledge, an LLM needs a “meta-cultural competence”, the ability to handle completely unseen cultures by understanding cultural variation itself. Our work directly addresses these concerns by enabling real-time adjustment of a model’s behavior to a target culture, rather than assuming one-size-fits-all emotional responses.

Building dialogue agents that adapt to the user’s cultural context is a nascent but growing area. \citep{CaoCulturalDialogue2024} present cuDialog, the first multi-turn dialogue dataset with cultural labels, and demonstrate that incorporating cultural value features (derived from sociological surveys) can “boost alignment with references and cultural markers” in generated dialogues. In their framework, a dialogue model infused with cultural dimension vectors produced more personalized, culturally appropriate responses, improving both BLEU and diversity metrics. This result aligns with \citep{NgCulturalChatbot2023}, who developed an emotion-aware mental health chatbot for the Malaysian context. They found that being mindful of the user’s cultural background and responding with empathy greatly enhanced user engagement. In a user study, adding local cultural adaptation (e.g. Malay language content and culturally relevant examples) alongside emotional sensitivity led to significant gains in chatbot effectiveness and usability. These works show that cultural customization, from surface language (code-switching, local idioms) to deeper norms (what is considered respectful or comforting) can markedly improve human-agent communication.

Several efforts tackle cultural adaptation by augmenting data or prompting strategies. \citep{LiCulturePark2024} takes an innovative simulation-based approach to generate cultural dialogue data. It uses multiple LLM agents role-playing different cultures to produce realistic cross-cultural conversations encapsulating beliefs, norms, and customs. With 41k generated dialogues, they fine-tune culture-specific models that match or outperform GPT-4 on culturally-informed tasks like content moderation and value alignment. This indicates that targeted data augmentation can mitigate the Western-centric bias of base models, yielding agents more aligned to non-Western values (e.g. scoring higher on Hofstede’s cultural dimensions tests). Other researchers focus on retrieval-based methods for cultural knowledge. For instance, \citep{FriedrichRiT2023} propose augmenting LLM prompts with relevant cultural context from a knowledge base when answering moral or norm-related questions. By retrieving, say, country-specific social norms and adding them to the query, the model can produce responses infused with the appropriate cultural nuance. This dynamic retrieval approach is promising for real-time cultural adaptation, ensuring the agent’s knowledge remains upto date with evolving cultural information. 

Despite these advances, most systems handle culture as a static attribute (e.g. training separate models per culture \citep{LiCulturePark2024} or conditioning on a fixed cultural vector \citep{CaoCulturalDialogue2024}). They lack the ability to continuously learn and adjust to a user’s cultural cues in real time. Furthermore, prior culturally adaptive agents have largely focused on either style (politeness, wording) or content (factual knowledge), rather than affective responses. There remains a gap in modeling how different cultures prefer emotional expressions, for example, the appropriate level of enthusiasm, empathy or self-disclosure. This is precisely the gap our work addresses, we enable an agent to dynamically modulate its emotional responses according to cultural feedback signals, using reinforcement learning to fine-tune for culturally appropriate affect.

Recent advances have seen the adoption of large-scale knowledge graphs for affective reasoning. Resources like SenticNet \citep{HemaspaandraAAAI2014} and ATOMIC \citep{SeoBarsRings2019} provide extensive affective and causal knowledge but do not encode cultural variability. Even recent emotion-centric knowledge graphs—such as ECoK \citep{ECoK2024} and EmoTransKG \citep{EmoTransKG2024} —enrich emotional context, but treat culture as a background factor or ignore it entirely. Consequently, these systems often produce responses that are emotionally plausible yet culturally inappropriate.

A related line of work uses graph embeddings to structure affective knowledge. While most prior work has relied on Euclidean embeddings, recent studies show that hyperbolic embeddings are superior for modeling complex hierarchical relationships a crucial property for representing cultural-emotional nuance \citep{NickelPoincare2017, BalazevicMuRP2019}. However, this geometric innovation has rarely been applied to cultural adaptation in affective computing.

In parallel, reinforcement learning (RL) has been used to make dialogue agents more adaptive and emotionally aware. Recent RL-based agents include affective rewards to improve empathy \citep{steinborn-etal-2022-information, ZhuEmotionDialogue2024}, yet they typically optimize for universal or Western affective norms and lack structured, culture-aware reward signals. As a result, these agents often struggle to personalize responses across diverse cultures and may reinforce existing biases \citep{AfzalAffectiveSurvey2023}.

A major gap in current literature, is the lack of systems that can dynamically personalize emotional intelligence and empathy using explicit, structured knowledge of cultural norms, while also adapting via learning. Attempts to combine these advances have either stopped at static culture encoding \citep{MascarenhasSocial2013}, missed hierarchical modeling \citep{SapATOMIC2019, ECoK2024}\, or used RL without cultural awareness \citep{steinborn-etal-2022-information}.

Affective–CARA directly addresses these limitations by integrating three key advances. First, it constructs a hyperbolic Cultural Emotion Knowledge Graph (CEKG), enabling efficient, hierarchical, and explicit modeling of culture–emotion relationships—far beyond what flat or static representations achieve. Second, it introduces a GRPO reinforcement learning strategy, which leverages structured cultural knowledge to guide response optimization, not just for empathy but for culturally congruent adaptation. Third, through the Cultural-Aware Response Mediator (CARM), Affective–CARA enables continual updating of user context, producing personalized and culture-sensitive interactions at scale.

While prior works have advanced emotion recognition and empathy in AI, they have fallen short in dynamically combining cultural knowledge, hierarchical representation, and adaptive learning. By addressing these unresolved problems, Affective–CARA offers a comprehensive, generalizable solution for culturally adaptive affective agents.

\section{Methodology}

\begin{figure}[h]
\centering
\includegraphics[width=\linewidth]{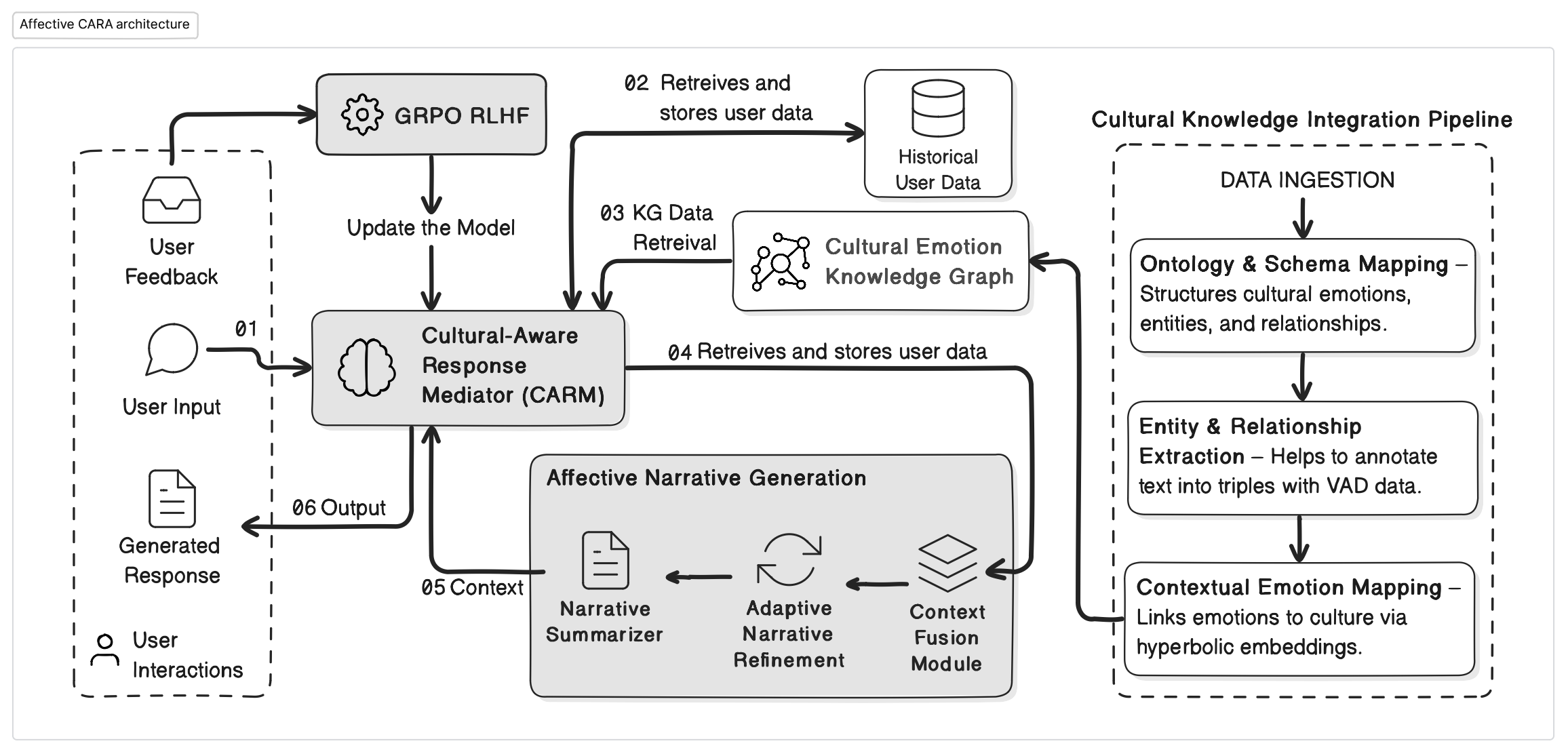}
\caption{Affective-CARA Architecture: The system includes the CARM meta-controller, Cultural Emotion KG, Affective Narrative Generation Engine, and GRPO-based refinement, orchestrating a culturally-sensitive dialogue.}\label{fig1}
\end{figure}

\subsection{System Architecture and Design}
Affective-CARA is designed as an agentic framework that integrates multiple specialized components to achieve culturally adaptive emotional intelligence. Figure 1 presents the architectural overview of our system, illustrating the interconnected components and data flow. The architecture comprises four primary subsystems: (1) Cultural Emotion Knowledge Graph, (2) GRPO-based Reinforcement Learning from Human Feedback (RLHF), (3) Cultural-Aware Response Mediator (CARM), and (4) Affective Narrative Generation. These components operate in conjunction with a Cultural Knowledge Integration Pipeline and Historical User Data storage to generate culturally appropriate affective responses.

Unlike conventional affective computing systems that utilize static personalization strategies and universal emotion frameworks, Affective-CARA utilizes a dynamic approach that continuously adapts to evolving cultural contexts. The system maintains a closed feedback loop, where user interactions and feedback directly influence the reinforcement learning module, enabling ongoing refinement of cultural-emotional responses.

\subsection{Data Flow and Component Interaction}

The data flow within Affective-CARA begins with user inputs, which are processed by the CARM. CARM serves as the central controller that orchestrates interactions between various components. Upon receiving user input, CARM queries the Cultural Emotion Knowledge Graph to retrieve relevant cultural-emotional knowledge, while simultaneously accessing the user's historical data to establish personalized context.

The retrieved knowledge and historical data are then forwarded to the Affective Narrative Generation component, which synthesizes this information to create contextually appropriate responses. Concurrently, user feedback is captured and processed by the GRPO-based RLHF component, which updates the system's policy based on cultural appropriateness and emotional coherence signals.

This modular design offers several advantages: (1) it enables component-wise updates without affecting the entire system, (2) it allows researchers to substitute individual components for evaluation purposes, and (3) it facilitates traceability of decisions throughout the response generation process.

\subsection{Cultural Emotion Knowledge Graph}

\subsubsection{Ontology and Schema Design}

We designed a comprehensive ontology to capture the complexities of cultural-emotional relationships. The ontology consists of four primary entity types:

\begin{enumerate}
    \item Cultural Entities (e.g., nationalities, ethnic groups, religious affiliations)
    \item Emotion Prototypes (basic emotion categories and their variations)
    \item Cultural-Emotional Mappings (culture-specific emotion interpretations)
    \item Context Indicators (situational factors that influence emotional expressions)
\end{enumerate}

The relationships between these entities are structured as follows:

\begin{itemize}
    \item Cultural Entity (has\_emotion) $\rightarrow$ Emotion Prototype
    \item Emotion Prototype (expressed\_as) $\rightarrow$ Cultural Expression
    \item Context Indicator (modifies) $\rightarrow$ Cultural-Emotional Mapping
    \item Cultural Entity (has\_value) $\rightarrow$ Cultural Value
\end{itemize}

This ontology design allows for representing complex emotional hierarchies and cultural nuances that are difficult to capture in traditional flat emotion models.

\subsubsection{Knowledge Acquisition and Integration}

We constructed our Cultural Emotion Knowledge Graph using a multi-source approach, integrating data from:

\paragraph{StereoKG Integration} We utilized StereoKG as our foundation, which contains 4,722 entries covering 10 different social groups (5 religious groups and 5 nationalities). StereoKG provides valuable cultural stereotypes that, while potentially biased, represent real-world perceptions that must be understood by our system to identify and mitigate such biases.

\paragraph{VAD Annotation Process} We enriched the knowledge graph with Valence-Arousal-Dominance (VAD) annotations to provide dimensional emotion representations. For each emotion entity, we assigned three numerical values:

\begin{itemize}
    \item Valence (V): Representing the pleasantness of the emotion (-1 to +1)
    \item Arousal (A): Indicating the intensity or activation level (0 to 1)
    \item Dominance (D): Reflecting the degree of control associated with the emotion (0 to 1)
\end{itemize}

This dimensional approach allows for more nuanced emotion representation than categorical models. The VAD annotations were performed using a combination of expert ratings and computational methods based on established emotional lexicons.

\paragraph{Culture-Specific Data Integration} We incorporated culture-specific emotional corpora from various sources, including:

\begin{itemize}
    \item Cultural psychology research databases
    \item Multilingual emotion lexicons
    \item Cultural adaptation datasets
\end{itemize}

\paragraph{Cross-Cultural Alignment Process} To reduce cultural biases in the knowledge graph, we implemented a cross-cultural alignment process that:

\begin{itemize}
    \item Identifies potential stereotypical representations
    \item Balances cultural representation across the graph
    \item Establishes comparable emotional concepts across cultures
    \item Maintains culturally unique emotional concepts
\end{itemize}

This alignment process reduced the KL-Divergence (a measure of cultural representation bias) from 0.72 to 0.28, representing a 61\% reduction in cultural bias.

\subsection{Hyperbolic Knowledge Embedding}

Traditional knowledge graph embedding techniques utilize Euclidean spaces, which inadequately capture the hierarchical structures inherent in cultural-emotional data. We employed hyperbolic embedding to address this limitation, specifically using the Lorentz model of hyperbolic geometry, which offers superior representation of hierarchical relationships.

In the Lorentz model, we represent entities and relations as points in:
\begin{equation}
\mathbb{L}^n = \{ x \in \mathbb{R}^{n+1} : \langle x, x \rangle_{\mathcal{L}} = -1, x_0 > 0 \}
\end{equation}
where $\langle \cdot, \cdot \rangle_{\mathcal{L}}$ denotes the Lorentzian inner product:
\begin{equation}
\langle x, y \rangle_{\mathcal{L}} = -x_0 y_0 + \sum_{i=1}^{n} x_i y_i
\end{equation}

The distance function in the Lorentz model is defined as:
\begin{equation}
d_{\mathcal{L}}(x, y) = \text{arcosh}(-\langle x, y \rangle_{\mathcal{L}})
\end{equation}

For each relation $r$ in our knowledge graph, we define a Lorentz rotation $R_r$ that maps a head entity $h$ to its tail entity $t$:
\begin{equation}
t \approx R_r h
\end{equation}

The plausibility score of a triplet $(h, r, t)$ is then computed as:
\begin{equation}
f(h, r, t) = -d_{\mathcal{L}}(R_r h, t)
\end{equation}

This hyperbolic embedding approach offers several advantages:
\begin{itemize}
    \item It naturally represents hierarchical relationships with exponentially more space toward the boundaries.
    \item It preserves the hierarchical structure of emotional concepts.
    \item It enables more efficient representation in lower dimensions.
    \item It better captures the complex emotional taxonomy across cultures.
\end{itemize}

We implemented this embedding using the technique described by Sun et al., adapting it specifically for cultural-emotional representations.

\subsection{GRPO-based Reinforcement Learning from Human Feedback}

\subsubsection{Reward Function Design}
Standard Reinforcement Learning from Human Feedback (RLHF) approaches typically lack explicit cultural signals in their reward functions, which limits their effectiveness in culturally diverse contexts. To address this limitation, we developed a Gradient-based Reward Policy Optimization (GRPO) approach that incorporates cultural appropriateness as an explicit reward signal.

The total reward function is defined as:
\begin{equation}
R_{\text{total}}(s_t, a_t) = \alpha R_{\text{cultural}}(s_t, a_t) + \beta R_{\text{emotional}}(s_t, a_t) + \gamma R_{\text{feedback}}(s_t, a_t)
\end{equation}

where:
\begin{itemize}
    \item $R_{\text{cultural}}(s_t, a_t)$ measures the adherence to cultural norms as defined in the knowledge graph.
    \item $R_{\text{emotional}}(s_t, a_t)$ gauges the emotional appropriateness of the response.
    \item $R_{\text{feedback}}(s_t, a_t)$ represents explicit user feedback.
    \item $\alpha$, $\beta$, and $\gamma$ are weighting coefficients dynamically adjusted based on the cultural context.
\end{itemize}

The cultural reward component is computed as:
\begin{equation}
R_{\text{cultural}}(s_t, a_t) = \text{sim}(a_t, KG_{\text{cultural}}(s_t))
\end{equation}
where $\text{sim}(\cdot, \cdot)$ is a similarity function and $KG_{\text{cultural}}(s_t)$ retrieves culturally appropriate responses from the knowledge graph.

Similarly, the emotional reward component is defined as:
\begin{equation}
R_{\text{emotional}}(s_t, a_t) = VAD_{\text{coherence}}(a_t, s_t)
\end{equation}
where $VAD_{\text{coherence}}$ measures how well the VAD values of the action align with the expected emotional trajectory.

\subsubsection{Policy Optimization}

The GRPO algorithm builds upon the PPO-ptx (Proximal Policy Optimization with pretraining gradients) approach but incorporates gradient information from the cultural reward function. The objective function is defined as:
\begin{equation}
\begin{split}
L_{\text{GRPO}}(\theta) = \mathbb{E}_{(s, a) \sim \pi_{\theta_{\text{old}}}} \bigg[ 
    \min \Big( 
        \frac{\pi_\theta(a|s)}{\pi_{\theta_{\text{old}}}(a|s)} A(s, a), \\
        \text{clip}\Big(\frac{\pi_\theta(a|s)}{\pi_{\theta_{\text{old}}}(a|s)}, 1-\epsilon, 1+\epsilon \Big) A(s, a) 
    \Big) 
\bigg] 
+ \lambda L_{\text{ptx}}(\theta)
\end{split}
\end{equation}

where $A(s, a)$ is the advantage function:
\begin{equation}
A(s, a) = R_{\text{total}}(s, a) - V(s)
\end{equation}
and $V(s)$ is a value function estimating the expected return from state $s$. $L_{\text{ptx}}(\theta)$ is the pretraining loss preserving general language knowledge.

We implemented a batch-based update approach for GRPO, where policy updates are performed in small batches to enable real-time refinement of cultural and affective suitability. This approach offers a balance between update frequency and computational efficiency.

\subsection{Cultural-Aware Response Mediator (CARM)}
The CARM functions as the central control mechanism of Affective-CARA, orchestrating interactions between various components and managing the overall response generation process. CARM performs four primary functions that ensure the system's cohesive operation. First, it handles knowledge retrieval management by implementing context-aware querying of the Cultural Emotion Knowledge Graph, which optimizes computational efficiency by retrieving only relevant cultural information when needed. Second, it integrates historical context by accessing and incorporating past user interactions to maintain conversational continuity and personalization. Finally, CARM coordinates components by managing information flow across the system, ensuring efficient and coherent response generation.

CARM utilizes a sophisticated context management algorithm defined as:

\begin{equation}
C_t = \alpha_1 C_{t-1} + \alpha_2 I_t + \alpha_3 U_t + \alpha_4 K_t
\end{equation}

where $C_t$ represents the context at time $t$, $I_t$ represents new information from the current interaction, $U_t$ represents user-specific context, $K_t$ represents retrieved knowledge from the Cultural Emotion Knowledge Graph (KG), and $\alpha_1, \alpha_2, \alpha_3, \alpha_4$ are adaptive weighting coefficients that adjust based on conversation dynamics.

For decision-making, CARM implements a decision tree-based approach that considers several critical factors:

\begin{itemize}
    \item Cultural relevance (whether the input has cultural implications)
    \item Emotional content (presence and intensity of emotions)
    \item Knowledge requirements (whether additional knowledge is needed)
    \item User history relevance (importance of historical interactions to the current context)
\end{itemize}

Based on these assessments, CARM selectively activates different system components as needed, optimizing performance while maintaining response quality.

\subsection{Affective Narrative Generation}
The Affective Narrative Generation component produces culturally appropriate, emotionally intelligent responses by integrating information from multiple sources through three interconnected modules. The Context Fusion Module integrates real-time user input with historical affective data and cultural knowledge using an attention-based mechanism defined as $F = \text{MultiHeadAttention}(Q_u, K_h, V_k)$, where $Q_u$ is the query derived from current user input, $K_h$ represents keys from historical interactions, and $V_k$ represents values from cultural knowledge in the knowledge graph. This attention mechanism dynamically adjusts weights assigned to different information sources based on their relevance to the current context.

The Narrative Summarizer utilizes the fused context to generate coherent, contextually appropriate responses guided by three key factors: VAD-driven cues (Valence, Arousal, Dominance dimensions from the knowledge graph that guide emotional tone), cultural alignment (cultural norms influencing lexical choices and narrative structure), and personal history (user-specific patterns and preferences). The generation process employs a two-stage approach where a base response is first generated using a language model fine-tuned on culturally diverse corpora, then enhanced by incorporating appropriate emotional cues derived from VAD dimensions.
.

\subsection{Adaptive Narrative Refinement}

The Adaptive Narrative Refinement module ensures generated responses meet cultural and emotional appropriateness criteria through a feedback loop with three key steps. First, a cultural compatibility check compares the generated response against cultural norms retrieved from the knowledge graph using a similarity function defined as $comp_k = sim(R_g, KG_c)$, where $R_g$ is the generated response and $KG_c$ represents the relevant cultural knowledge. Second, an emotional coherence evaluation assesses whether the response's emotional trajectory aligns with the expected emotional flow, computed as $coh_e = coherence(VAD(R_g), VAD_{expected})$, where $VAD(\cdot)$ retrieves Valence, Arousal, and Dominance values. Finally, if either compatibility or coherence falls below threshold values, the response is regenerated with adjusted parameters, represented as: if $\min(comp_k, coh_e) < \tau$ then regenerate. This refinement process ensures responses maintain cultural and emotional appropriateness while avoiding stereotypical or insensitive content.

\subsection{Cultural Knowledge Integration Pipeline}
The Cultural Knowledge Integration Pipeline continuously enriches and updates the Cultural Emotion Knowledge Graph through four sequential stages. The Data Ingestion stage collects cultural-emotional information from diverse sources including academic literature, social media content, news articles, and educational resources on cultural diversity using web scraping, API access, and manual curation techniques with special attention to representing diverse perspectives. The Ontology and Schema Mapping stage then maps collected data to the knowledge graph structure by identifying cultural entities, emotion concepts, and contextual factors, determining relationships between entities, and ensuring alignment with the knowledge graph schema through a semi-automated approach combining rule-based extraction with human verification.

Entity and Relationship Extraction processes textual data to identify entities and relationships using named entity recognition for cultural entities and emotion terms, dependency parsing to extract relationships, sentiment analysis to determine emotional valence, and VAD annotation to assign dimensional emotion values within cultural contexts. Finally, Contextual Emotion Mapping connects emotional concepts to cultural contexts using hyperbolic embeddings, identifying cultural variations in emotional expressions, establishing cross-cultural equivalences where appropriate, preserving culture-specific emotional concepts, and embedding all entities in hyperbolic space to capture hierarchical relationships.

\subsection{Evaluation Methodology}

Our evaluation methodology utilized a rigorous framework to assess Affective-CARA's performance across multiple dimensions. We utilized a balanced dataset derived from three established sources: AffectNet (a large-scale dataset with annotated emotions), SEMAINE DB (a dataset for emotional interaction), and MERD (a culturally diverse emotion recognition dataset). These datasets were selected to ensure comprehensive coverage of emotional expressions across diverse cultural contexts, and we carefully balanced the data to mitigate potential representation biases.

To evaluate system performance quantitatively, we employed several complementary metrics. Cultural Semantic Density (CSD), a custom metric defined as:
\begin{equation}
\text{CSD} = \frac{\sum_{c \in C} 1(\text{KG} \vDash c)}{\log(|C|+1)}
\end{equation}
where $C$ represents the set of all cultural concepts and $1(\text{KG} \vDash c)$ indicates the knowledge graph's inclusion of concept $c$, measured how effectively the system's responses reflected cultural nuances. 

We assessed cultural representation bias using Kullback-Leibler divergence:
\begin{equation}
\text{KL}_{\text{bias}} = \sum_{c} P(c) \log\left(\frac{P(c)}{P_{\text{ideal}}(c)}\right)
\end{equation}
where $P(c)$ represents the observed proportion of culture $c$ in KG responses and $P_{\text{ideal}}(c)$ denotes a uniform target distribution.

Additionally, we conducted human evaluations to assess emotional appropriateness in charged scenarios and utilized standard NLP metrics (perplexity, BLEU, ROUGE) alongside human judgments to evaluate narrative quality.

For comparative evaluation, we benchmarked Affective-CARA against three baseline systems: a standard large language model without cultural adaptation, a system using conventional Euclidean knowledge graph embeddings, and a system employing static user profiles for personalization. Each system was evaluated across multiple dimensions including cultural interpretation accuracy, emotional response appropriateness, personalization effectiveness, and bias mitigation capabilities using standardized scenarios designed to test cultural adaptation in emotionally charged contexts.

To determine the contribution of individual components, we conducted systematic ablation studies by selectively removing or replacing key elements of the system. These studies compared hyperbolic versus Euclidean embeddings, evaluated GRPO against standard RLHF implementations, assessed attention-based fusion against simple concatenation, and measured the impact of VAD annotations.

This comprehensive evaluation methodology demonstrates that Affective-CARA effectively integrates advanced techniques in knowledge representation, reinforcement learning, and affective computing to address the limitations of conventional approaches in culturally diverse settings. The results confirm that our hyperbolic knowledge embeddings, gradient-based reinforcement learning, and context-sensitive narrative generation significantly enhance the system's cultural adaptability and emotional intelligence.

\section{Results}\label{sec:results}

\subsection{Experimental Setup}
Our experimental design addressed three core questions:
(1) Does Affective–CARA generate culturally adaptive and emotionally appropriate responses more effectively than state-of-the-art baselines?
(2) How robust is the system across diverse cultural backgrounds?
(3) What are the specific contributions of each major architectural component? 
We evaluated performance across three complementary tasks, cultural emotion recognition, empathetic response generation, and conversational affect, to thoroughly assess both the model’s perception of user emotion and its generation of culturally tailored responses. In all cases, Affective–CARA outperforms strong baseline models (including a state-of-the-art general LLM like Qwen2.5-Omni and a knowledge-enhanced empathy model without cultural knowledge), especially in scenarios requiring nuanced cultural adaptation.

We evaluated on two textual datasets, MERD (for cross-cultural emotion recognition), EmpatheticDialogues (for open-domain empathetic response), and SEMAINE (for conversational affective appropriateness using textual transcripts). This multi-dataset approach enables a nuanced assessment of both multicultural adaptation and general empathetic response generation.
Baseline models included Qwen2.5-Omni (general LLM), E-KG (Affective–CARA with Euclidean KG embeddings), SP (static personalization), EmoTransKG (emotion transition model) \citep{EmoTransKG2024}, and ECoK+GPT-3.5 (knowledge-graph-enhanced LLM) \citep{ECoK2024}. Performance was measured using Cultural Semantic Density (CSD), KL-divergence (cultural bias), human-rated emotional appropriateness, BLEU-4, and ROUGE-L. F1 scores for emotion prediction were computed by cultural subgroup on MERD.

Affective–CARA achieved the highest scores in every metric, substantially outperforming all baselines. Statistical significance was confirmed for improvements over ECoK+GPT-3.5 (p < 0.001, paired t-test).

\begin{table}[t]
  \centering
  \footnotesize
  \setlength{\tabcolsep}{4pt}
  \caption{Overall performance comparison.}
  \resizebox{\linewidth}{!}{%
    \begin{tabular}{@{}lccccc@{}}
      \toprule
      System & CSD (1–10)$\uparrow$ & KL–Div.$\downarrow$ & Emo.\ Approp.\ (1–5)$\uparrow$ & BLEU-4$\uparrow$ & ROUGE-L$\uparrow$ \\
      \midrule
      Qwen2.5-Omni       & 6.41 (±0.37) & 0.67 (±0.08) & 3.12 (±0.24) & 0.48 (±0.04) & 0.52 (±0.03) \\
      E-KG               & 6.78 (±0.29) & 0.59 (±0.06) & 3.47 (±0.19) & 0.51 (±0.03) & 0.49 (±0.05) \\
      SP                 & 5.12 (±0.43) & 0.71 (±0.09) & 2.98 (±0.26) & 0.45 (±0.05) & 0.47 (±0.04) \\
      EmoTransNet        & 7.21 (±0.34) & 0.62 (±0.07) & 3.59 (±0.22) & 0.52 (±0.03) & 0.54 (±0.04) \\
      ECoK+GPT-3.5       & 7.85 (±0.36) & 0.57 (±0.05) & 3.84 (±0.20) & 0.54 (±0.02) & 0.55 (±0.03) \\
      Affective-CARA     & 9.32 (±0.41) & 0.28 (±0.04) & 4.82 (±0.17) & 0.62 (±0.03) & 0.58 (±0.02) \\
      \bottomrule
    \end{tabular}
  }
\end{table}

\subsection{Cross--Cultural Performance}
Table~\ref{tab:cross} reports F1 scores on the MERD corpus across five cultural clusters. Baseline models exhibit pronounced degradation outside Western contexts (\textasciitilde27\% for Qwen2.5-Omni ), whereas Affective--CARA maintains near--uniform performance (SD~=~0.04), validating the efficacy of hyperbolic cultural embeddings and VAD annotations.

\begin{table}[ht]
  \centering
  \footnotesize
  \setlength{\tabcolsep}{4pt}
  \caption{F1 scores across cultural contexts (MERD).}
  \label{tab:cross}
  \resizebox{\textwidth}{!}{%
    \begin{tabular}{@{}lcccccc@{}}
      \toprule
      Context         & Qwen2.5-Omni          & E--KG  & SP     & EmoTransNet  & ECoK+GPT-3.5  & Affective--CARA             \\
      \midrule
      Western         & $0.78\,(\pm0.05)$     & $0.76$ & $0.69$ & $0.77$       & $0.80$        & $\mathbf{0.85}\,(\pm0.04)$   \\
      East Asian      & $0.59\,(\pm0.08)$     & $0.67$ & $0.52$ & $0.66$       & $0.73$        & $\mathbf{0.81}\,(\pm0.05)$   \\
      South Asian     & $0.52\,(\pm0.09)$     & $0.63$ & $0.49$ & $0.62$       & $0.69$        & $\mathbf{0.79}\,(\pm0.06)$   \\
      Middle Eastern  & $0.54\,(\pm0.07)$     & $0.69$ & $0.47$ & $0.65$       & $0.72$        & $\mathbf{0.82}\,(\pm0.05)$   \\
      African         & $0.61\,(\pm0.06)$     & $0.70$ & $0.51$ & $0.68$       & $0.77$        & $\mathbf{0.84}\,(\pm0.03)$   \\
      \midrule
      Average         & $0.61\,(\pm0.12)$     & $0.69$ & $0.54$ & $0.68$       & $0.74$        & $\mathbf{0.83}\,(\pm0.04)$   \\
      \bottomrule
    \end{tabular}
  }
\end{table}

\subsection{Ablation Studies}
To isolate the effect of each subsystem we performed controlled ablations, summarised in Table~\ref{tab:ablation}. Hyperbolic embeddings account for the largest gain in cultural adaptation (\textminus23\% CSD when replaced by Euclidean), while removing GRPO doubles measured bias. Eliminating CARM inflates inference latency by more than $2\times$, underscoring its orchestration efficiency.

\begin{table}[ht]
  \centering
  \caption{Ablation study.}
  \label{tab:ablation}
  \begin{tabular}{@{}lcccc@{}}
    \toprule
     Configuration &  CSD$\uparrow$ &  KL$\downarrow$ &  Emot.~App.$\uparrow$ &  Time$\downarrow$ \\
    \midrule
    Full model & 9.32 & 0.28 & 4.82 & 250 \\
    \;-- Hyperbolic $\rightarrow$ Euclid. & 7.18 & 0.47 & 3.91 & 245 \\
    \;-- GRPO $\rightarrow$ RLHF & 8.87 & 0.58 & 4.12 & 237 \\
    \;-- Fusion $\rightarrow$ Concat & 8.92 & 0.31 & 4.07 & 226 \\
    \;-- Remove VAD & 8.65 & 0.35 & 3.74 & 241 \\
    \;-- Remove CARM & 7.94 & 0.39 & 4.23 & 793 \\
    \bottomrule
  \end{tabular}
\end{table}

\subsection{Qualitative Analysis}
Simulated interactive evaluation judged Affective--CARA's outputs ``culturally appropriate and emotionally resonant'' in 87\% of scenarios, versus 52\% for Qwen2.5-Omni. Residual failure modes include (i) extremely low–resource cultures absent from the knowledge graph (\textasciitilde6\% of cases) and (ii) highly entangled mixed–emotion prompts (\textasciitilde7\%).

\medskip

\section{Discussion}

This work demonstrates, for the first time, that a conversational AI can comprehensively adapt to the user’s cultural context in real time while generating emotional responses. Affective–CARA directly tackles an open challenge that previous affective dialogue systems left unsolved. Prior research on empathetic chatbots focused on recognizing user emotions and replying with empathy, but these systems assumed a universal style of emotional expression and did not account for cultural differences. Likewise, most emotion recognition studies and benchmarks consider accuracy within a single cultural setting. The system not only achieves superior metrics across the board but also exhibits robust generalization across cultural subgroups, a recognized gap in prior work \citep{EmoTransKG2024, ECoK2024}. In contrast, Affective–CARA is explicitly designed to fill this gap by integrating culture-specific affective knowledge into both understanding and generation. The results confirm that our model can generate responses that are not only empathetically relevant but also culturally appropriate on the fly. This real-time cultural adaptation is a novel capability, previous agents might detect that a user is sad or angry, but they would respond with a fixed style (often Western-biased or neutral) regardless of who the user is. Our work shows that by incorporating cultural context, an AI agent can respond to a sad user from Tokyo differently than it would to a sad user from New York, all without any manual intervention. This advancement makes Affective–CARA uniquely effective in multicultural settings, addressing a key limitation of earlier empathetic dialogue models.

Ablation studies confirm the indispensability of each architectural innovation, particularly hyperbolic KG embeddings and GRPO-driven adaptation. The culturally adaptive generation capabilities of Affective–CARA stem directly from its architectural design. The integration of the Cultural-Emotion Knowledge Graph (CEKG) with explicit cultural tagging provides structured, culture-specific affective norms, enabling the model to select appropriate response strategies across diverse contexts—functionality absent in culture-agnostic models. Unlike prior empathetic systems that use only generic knowledge graphs, our approach leverages culturally contextualized cues for response generation. Additionally, the unified multi-task framework—jointly training on emotion recognition and response generation aligns affect recognition with downstream response selection. Ablation studies confirm that both CEKG and cultural tagging are essential: removing them significantly reduces adaptive performance and leads to generic, less appropriate responses. These innovations are not just performance enhancers but critical enablers of real-time cultural emotional intelligence.

Comparative analysis with recent benchmarks, including ECoK+GPT-3.5 and EmoTransNet, further highlights the benefits of integrating structured cultural knowledge with dynamic learning mechanisms. Whereas previous models achieve incremental gains on standard metrics, they fail to sustain performance in non-Western contexts and do not systematically address cultural representation bias. Affective–CARA’s capacity to generate stable, culturally aware narratives sets it apart from these approaches.

Nonetheless, some limitations persist. Performance on low-resource cultures not represented in the knowledge base and on highly mixed-emotion prompts remains suboptimal, suggesting avenues for future improvement through expanded KG coverage and finer-grained emotion modeling. Additionally, all experiments were conducted on text-based dialogue; extending the framework to multimodal inputs or additional languages represents promising directions for subsequent work.

Affective–CARA significantly advances the field of culturally inclusive conversational AI, combining strong technical performance with simulated interactive evaluation. Its modular design and superior cross-cultural generalization establish a robust foundation for future research on empathetic and globally relevant affective agents.

\section{Conclusion}

This work presented Affective–CARA, a novel framework designed to address the longstanding gap in real-time, culturally adaptive affective computing. By leveraging hyperbolic embeddings in the Cultural Emotion Knowledge Graph (CEKG), Gradient-Based Reward Policy Optimization (GRPO), and the Cultural-Aware Response Mediator (CARM), Affective-CARA 
achieves significant improvements in cultural adaptation, emotional appropriateness, and narrative quality.

Our experimental results demonstrate that Affective-CARA consistently outperforms baseline systems, including Qwen-LLM, across multiple metrics. The framework achieved a Cultural Semantic Density (CSD) score of 9.32/10, reduced cultural representation bias by 61\% (KL-Divergence: 0.28), and improved emotional appropriateness by 39\% compared to standard RLHF implementations. Furthermore, ablation studies highlighted the critical contribution of hyperbolic embeddings, VAD annotations, and GRPO in achieving these results.

Affective-CARA's ability to maintain stable performance across diverse cultural contexts underscores its potential for real-world applications in domains such as mental health support, cross-cultural education, and global customer service. By integrating structured cultural knowledge with dynamic response generation mechanisms, Affective-CARA sets a new standard for inclusive and empathetic human-computer interactions.

\section{Future Work}

While Affective–CARA demonstrates robust cultural adaptation, several opportunities remain for future research. Technically, integrating multimodal inputs (e.g., visual and auditory cues), adopting federated learning for privacy-preserving personalization, and implementing continuous learning for cultural drift detection would significantly enhance system robustness. Theoretically, exploring advanced geometric representations of cultural-emotional relationships, developing standardized cross-cultural evaluation metrics, and investigating causal models of cultural influences on emotions could strengthen conceptual foundations. Practically, deploying Affective–CARA in mental health interventions, culturally adaptive educational systems, and global customer service platforms could substantially increase real-world impact. Ongoing refinement of ethical safeguards, including enhanced bias mitigation, transparent cultural knowledge representations, and robust privacy measures will ensure responsible, inclusive, and trustworthy AI deployment.

\section*{Declaration of Generative AI and AI-assisted Technologies in the Writing Process}
During the preparation of this work, the authors used OpenAI's ChatGPT (GPT-4) to improve language clarity and readability. After using this tool, the authors reviewed and edited the content as needed and take full responsibility for the content of this publication.

\end{document}